\documentclass[%
reprint,
superscriptaddress,
amsmath,amssymb,
aps,
]{revtex4-1}

\usepackage{gensymb}
\usepackage{graphicx}% Include figure files
\usepackage{dcolumn}% Align table columns on decimal point
\usepackage{bm}% bold math
\usepackage{color}
\usepackage{braket}
\usepackage[normalem]{ulem}
\usepackage{caption}
\usepackage{subcaption}
\usepackage{bbold}
\newcommand\ai{{\it ab initio}}
\newcommand\ea{{\it et al.}}

\newcommand\ie{{\it i.e.}}

\newcommand\rr{\hat{\bm{r}}}

\begin{document}
	
	\preprint{APS/123-QED}
	
	\title{Light--absorbed orbital angular momentum in the linear response regime}
	
	\author{Philippe Scheid}
	\email{philippe.scheid@univ-lorraine.fr}
	\affiliation{
    Universit\'e de Lorraine, LPCT, CNRS, UMR 7019, BP 70239,  54506 Vandoeuvre-l\`es-Nancy Cedex, France
	}
	\affiliation{
	Universit\'e de Lorraine, IJL, CNRS, UMR 7198, BP 70239, 54000 Nancy Cedex, France
	}

	\author{Stéphane Mangin}
	\affiliation{
	Universit\'e de Lorraine, IJL, CNRS, UMR 7198, BP 70239, 54000 Nancy Cedex, France
	}

	\author{Sébastien Lebègue}
	\affiliation{
	Universit\'e de Lorraine, LPCT, CNRS, UMR 7019, BP 70239,  54506 Vandoeuvre-l\`es-Nancy Cedex, France
	}

	\date{\today}% It is always \today, today,
	%  but any date may be explicitly specified
	
	\begin{abstract}
		In exploring the light-induced dynamics within the linear response regime, this study investigates the induced orbital angular momentum on a wide variety of electronic structures. We derive a general expression for the torque induced by light on different electronic systems based on their characteristic dielectric tensor. We demonstrate that this phenomenon diverges from the inverse Faraday effect as it produces an orbital magnetization persistent post-illumination. Indeed, our results reveal that, while isotropic non-dissipative materials do not absorb orbital angular momentum from circularly polarized light, any symmetry-breaking arrangement of matter, be it spatial or temporal, introduces novel channels for the absorption of orbital angular momentum, or magnetization. Most notably, in dissipative materials, circularly polarized light imparts a torque corresponding to a change in orbital angular momentum of $\hbar$ per absorbed photon. The potential of these mechanisms to drive helicity-dependent magnetic phenomena paves the way for a deeper understanding of light-matter interactions. Notably, the application of pump-probe techniques in tandem with our findings allows experimentalists to quantitatively assess the amount of orbital angular momentum transferred to electrons in matter, thus hopefully enhancing our ability to steer ultrafast light-induced magnetization dynamics.
	\end{abstract}
	
	\pacs{Valid PACS appear here}% PACS, the Physics and Astronomy
	% Classification Scheme.
	%\keywords{Suggested keywords}%Use showkeys class option if keyword
	%display desired
	\maketitle

	%%%%%%%%%%%
	%% INTRO
	%%%%%%%%%%%
	\section{Introduction}
	
	With the study of phenomena such as the birefringence, the photoelectric effect and, more recently, a plethora of ultrafast light-induced dynamics, the interaction between light and matter has remained a very fertile ground for physics throughout the past centuries. Most notably, the explanation of the photoelectric effect by Einstein\cite{einstein_uber_1905} not only shed light on the mechanisms of absorption of energy carried by light, but also on the quantum nature of light itself.\\
	Conversely, mechanisms by which the angular momentum of light and matter are interacting with each other have remained a challenging topic. Michael Faraday pioneered this field in 1845 with the discovery that linearly polarized light rotates while propagating through a piece of heavy glass subjected to an external magnetic field. More than one century later, Pitaevskii\cite{pitaevskii_electric_1961} theoretically showed that circularly polarized light induces a magnetic moment proportional to its intensity in non-dissipative media, \ie\ that light can also change the magnetic state of matter. He naturally coined this phenomena ``inverse Faraday effect" (IFE). Only a few years later, experimental confirmation of a static magnetization solely induced by the light was given\cite{van_der_ziel_optically-induced_1965}. With the advance of femtosecond ultra-intense light sources and the discovery of the all-optical helicity-dependent switching effect\cite{stanciu_all-optical_2007, mangin_engineered_2014, lambert_all-optical_2014, scheid_light_2022}, where the polarization of the light can be used to steer the magnetic state of metallic magnets in a deterministic way, the interest for the IFE has even been further invigorated. However, due to the continuum of states available at the Fermi energy and above it, these compounds are highly absorptive and the traditional IFE could no longer be invoked. To this end, Battiato \ea\cite{battiato_beyond_2012, battiato_quantum_2014} developed a generalized framework to compute it in any type of material.\\
	Before Pitaevskii, and differently from an induced magnetic moment as in the IFE, Beth\cite{beth_mechanical_1936} experimentally showed that, while light travels in a transparent birefringent compound, and its polarization evolves from linear to circular, a torque is exerted on the material. He assigned this applied torque to the fact that, for light to be able to change its elipticity, it has to gain or lose angular momentum, which is thus lost or gained by the material such that the angular momentum remains conserved. More recently, the spin counterpart of this angular-momentum absorption has been theoretically exhibited in dissipative magnetic materials\cite{scheid_ab_2019, scheid_ab_2021}, and is enabled by the presence of spin-orbit coupling.
	%\cite{Canaguier-Durand2013, Canaguier-Durand2013a}\\
	Here, building on the initial work of Beth\cite{beth_mechanical_1936} and Scheid \ea\cite{scheid_ab_2019, scheid_light_2022}, we derive the general expression of the torque induced by light on the electronic system, directly as a function of a measurable quantity: the dielectric tensor. Doing so, we readily express the density of absorbed orbital angular momentum in a wide variety of materials, and, using experimentally measured dielectric tensors, we provide exact values for the latter.\\
	Notably, we show that the absorbed orbital angular momentum cancels in isotropic non-dissipative materials, even for circularly polarized light, thus contrasting with the IFE. Additionally, we find that a linearly polarized electric field induces a second harmonic generation of orbital angular momentum in non-absorptive but magnetic materials. As this effects arises directly from the same broken symmetries as the Faraday effect, it thus appears to be the actual counterpart of the latter.
	In contrast, we show that in dissipative materials, the absorption of circularly polarized light induces a torque on the electronic system, which direction is opposite for both polarization of the light and that corresponds to a change of orbital angular momentum of $\hbar$ per absorbed photon.
	Lastly, we examine non-isotropic materials, which present additional channels for the absorption of orbital angular momentum, even when they are not dissipative, \ie\ do not absorb energy.

	\section{Theory}
	
	\subsection{Light-induced torque and angular momentum as a function of linear response tensors}
	
	We are interested in the calculation of the torque, $\braket{\bm{\tau}}_{\text{ind}}(t)$, induced by the interaction of the light with the electronic system. We begin by showing that only the light--induced variation of the expectation value of the position operator, $\hat{\bm{r}}$, acting on any arbitrary electronic subspace, and which we write $\delta \braket{\hat{\bm{r}}}_{\text{ind}}(t) = \braket{\hat{\bm{r}}}(t) - \braket{\hat{\bm{r}}}_0$, is needed to assess both of these quantities. Here the brackets stand for the expectation value, which are performed on the light-induced time-evolving states, except when written as $\braket{}_0$ in which case it is performed on the ground state. $\braket{\hat{\bm{r}}}_0$ therefore is the ground state position of the electronic center of mass, static in the frame of the laboratory. Note that we ignore microscopic variations, \ie\ we consider a macroscopic average, which here is also performed on a homogeneous medium such that we can neglect any spatial dependence of the average.\\
	In this framework, the torque operator writes as:
	
	\begin{equation}
		\begin{split} 
			\hat{\bm{\Gamma}} & = \sum_i \left(\hat{\bm{r}}_i - \braket{\hat{\bm{r}}_i}_0\right) \times -e \bm{E} (t)\\
		\end{split}
	\end{equation}

	where $i$ is indexing the electronic subspaces, $-e$ is the charge of an electron and $\bm{E}(t)$ is the electric field of the light, such that the light-induced torque finally writes as:
	
	\begin{equation}
		\braket{\bm{\Gamma}}_\text{ind}(t) = \delta \braket{\hat{\bm{r}}_\text{tot}}_{\text{ind}}(t) \times -e \bm{E} (t)
		\label{eq:torque_rtot}
	\end{equation}

	where $\hat{\bm{r}}_\text{tot} = \sum_i \hat{\bm{r}}_i$, $i$ is indexing the electronic subspaces, \\
	Eq. \ref{eq:torque_rtot} can be further simplified on the account that, due to the symmetry of the wavefunction:
	
	\begin{widetext}
		\begin{equation} \label{eq:scaling}
			\begin{split} 
				\bra{\Psi(t)} \hat{\bm{r}}_i \ket{\Psi(t)} & = \int \Psi^*(\rr_1, \dots, \rr_i, \dots, \rr_N) \rr_i \Psi(\rr_1, \dots, \rr_i, \dots, \rr_N) d\rr_1 \dots d\rr_N\\
				& = \int \Psi^*(\rr_i, \rr_1, \dots \rr_N) \rr_i \Psi(\rr_i, \rr_1, \dots \rr_N) d\rr_1 \dots d\rr_N\\
				& = \int \Psi^*(\rr_1, \dots, \rr_N) \rr_1 \Psi(\rr_1, \dots, \rr_N) d\rr_1 \dots d\rr_N
			\end{split}
		\end{equation}
	\end{widetext}

	such that we have $\bra{\Psi(t)} \hat{\bm{r}}_i \ket{\Psi(t)} = \bra{\Psi(t)} \hat{\bm{r}}_1 \ket{\Psi(t)}$, \ie\ the expectation value of an operator acting on any subspace is the same, which simply translates the indistinguishability of identical particles. Consequently, $\braket{\rr_\text{tot}} (t) = N_e \braket{\rr} (t)$ where $N_e$ is the total number of electrons. Eq. \ref{eq:torque_rtot} can therefore equally be written in terms of the light-induced torque density, $\braket{\hat{\bm{\tau}}}_\text{ind}(t) = \frac{\braket{\bm{\Gamma}}_\text{ind}(t)}{V}$, where $V$ is the volume occupied by the material, and the electronic density, $\rho_e = \frac{N_e}{V}$, such as:
	
	\begin{equation}\label{eq:torque_ind_density}
		\braket{\bm{\tau}}_\text{ind}(t) = \rho_e \delta \braket{\hat{\bm{r}}}_\text{ind}(t) \times -e \bm{E} (t)
	\end{equation}

	Hence, when considering the light-induced polarization density one gets:
	
	\begin{equation}\label{eq:lin_resp}
		\begin{split}
			\braket{\hat{\bm{P}}}_\text{ind}(t) & = -e \rho_e \delta \braket{\rr}_\text{ind} (t)\\
				& = \varepsilon_0 \int_{-\infty}^{\infty} dt' \chi_{\bm{p}}(t-t') \bm{E}(t'),
 		\end{split}
	\end{equation}

	where the passage from the first to the second line is true within the framework of linear response. From Eq. \ref{eq:lin_resp}, the light-induced variation of the electronic center of mass finally expresses as:
	
	\begin{equation}\label{eq:r_ind}
		\begin{split}
			\delta \braket{\rr}_\text{ind} (t) & = \frac{\varepsilon_0}{-e \rho_e} \int_{-\infty}^{\infty} dt' \chi_{\bm{p}}(t-t') \bm{E}(t')\\
			& = \frac{\varepsilon_0}{-e \rho_e} \frac{1}{2 \pi} \int_{-\infty}^{\infty} d\omega \tilde{\chi}_{\bm{p}}(\omega) \tilde{\bm{E}}(\omega) e^{-i \omega t}.
		\end{split}
	\end{equation}

	Here $\tilde{\chi}_{\bm{p}}(\omega)$ and $\tilde{\bm{E}}(\omega)$ respectively are the Fourier transform of the polarizability tensor and of the electric field of the light. Indeed, we now interest ourselves in the spectrum of $\braket{\hat{\bm{\tau}}}_\text{ind}$ which, given by its obvious dependence on the polarizability and dielectric tensors are expected to vary with the frequency of the incident electromagnetic field, as well as their polarization and direction with respect to the crystallographic axis.\\
	To investigate such a dependence, we now consider the incident light to be a monochromatic planewave, which electric field can thus be written as:
	
	\begin{equation}
		\bm{E}(t) = E_0 \Re(e^{i \Omega t}\bm{u}),
	\end{equation}

	with $\bm{u}$ a Jones vector from which all the polarizations and propagation directions can be expressed and $E_0$ the amplitude of the electric field. It's Fourier transform expresses as:
	
	\begin{equation}\label{eq:E_ext_FT}
		\tilde{\bm{E}}(\omega) = \pi E_0 \left( \delta (\Omega + \omega) \bm{u} + \delta (\Omega - \omega) \bm{u}^* \right) 
	\end{equation}
	
	Using Eq. \ref{eq:torque_ind_density} and \ref{eq:E_ext_FT}, we can now compute the light-induced torque $\braket{\hat{\bm{\tau}}}_\text{ind}(t)$ exerted by arbitrarily polarized incident planewaves. Indeed, expressing Eq. \ref{eq:torque_ind_density} as a function of $\tilde{\chi}_{\bm{p}}(\omega)$ and $\tilde{\bm{E}}(\omega)$ yields:
	
	\begin{widetext}
	\begin{equation}\label{}
		\braket{\hat{\bm{\tau}}}_\text{ind} (t) = \frac{\varepsilon_0}{(2 \pi)^2} \int_{-\infty}^{+\infty} d\omega \int_{-\infty}^{+\infty} d\omega' \left( \tilde{\chi}_{\bm{p}}(\omega) \cdot \tilde{\bm{E}}(\omega) \right) \times \tilde{\bm{E}}(\omega') e^{- i (\omega + \omega') t}
	\end{equation}
	\end{widetext}

	which, upon injection of Eq. \ref{eq:E_ext_FT} gives:
	
	\begin{equation}\label{eq:torque}
		\begin{split}
			\braket{\hat{\bm{\tau}}}_\text{ind} (t) &= \frac{1}{2} \varepsilon_0 E^2_0 \big[
			\mathfrak{Re}\left[e^{-i 2 \Omega t} \left(\tilde{\chi}_{\bm{p}}(\Omega)\cdot \bm{u}^*\right) \times \bm{u}^*\right]\\
				& + \mathfrak{Re}\left[\left(\tilde{\chi}_{\bm{p}}(\Omega) \cdot \bm{u}^*\right) \times \bm{u} \right] \big].
		\end{split}
	\end{equation}

	Both of the latter quantities can equally be written in terms of the transversal dielectric response by considering the relation:
	
	\begin{equation}
		\varepsilon(\Omega) = \mathbb{1} + \chi_{\bm{p}}(\Omega),
	\end{equation}

	such that we finally obtain:
	
	\begin{equation}
		\begin{split}
			\braket{\hat{\bm{\tau}}}_\text{ind} (t) &= \frac{1}{2} \varepsilon_0 E^2_0 \big[
			\mathfrak{Re}\left[e^{-i 2 \Omega t} \left(\tilde{\varepsilon}(\Omega)\cdot \bm{u}^*\right) \times \bm{u}^*\right]\\
			& + \mathfrak{Re}\left[\left(\tilde{\varepsilon}(\Omega) \cdot \bm{u}^*\right) \times \bm{u} \right] \big],
		\end{split}
	\end{equation}

	or, equivalently in term of the intensity, I:
	
	\begin{equation}\label{eq:torque}
	\begin{split}
		\braket{\hat{\bm{\tau}}}_\text{ind} (t) &= \frac{1}{c} I \big[
		\mathfrak{Re}\left[e^{-i 2 \Omega t} \left(\tilde{\varepsilon}(\Omega)\cdot \bm{u}^*\right) \times \bm{u}^*\right]\\
		& + \mathfrak{Re}\left[\left(\tilde{\varepsilon}(\Omega) \cdot \bm{u}^*\right) \times \bm{u} \right] \big],
	\end{split}
	\end{equation}

	with $c$ the speed of light.

	Using Eq. \ref{eq:torque}, one can thus compute the total density of orbital angular momentum absorbed by any compound, as long as the intensity and $\varepsilon$ are known as a function of time.
	
	\subsection{Case of a planewave propagating along the $z$ axis in different types of media}\label{sec:application}
	
	We now proceed to the evaluation of the amplitude of the light-induced torque by using, as much as possible, experimentally measured values of the dielectric function. In their ground state, this quantity has been measured for a wide range of materials, such that in what follows $\varepsilon$ is considered time-independent. By considering this, instead of assessing the induced torque per unit of intensity, we can compute the induced orbital angular momentum per unit of fluence. Through the gyroscopic ratio, $\gamma$, we can thus compute the absorbed magnetization, $\bm{m}_\text{abs}$, per unit of fluence such as:
	
	\begin{equation}\label{eq:torque_F}
		\begin{split}
			\frac{d \bm{m}_\text{abs}}{d F} &= \frac{\gamma}{c} \big[
			\mathfrak{Re}\left[e^{-i 2 \Omega t} \left(\tilde{\varepsilon}(\Omega)\cdot \bm{u}^*\right) \times \bm{u}^*\right]\\
			& + \mathfrak{Re}\left[\left(\tilde{\varepsilon}(\Omega) \cdot \bm{u}^*\right) \times \bm{u} \right] \big].
		\end{split}
	\end{equation}
	
	Here we note the key difference between this induced torque, resulting in an induced magnetization per unit of fluence, and the induced magnetization per unit of intensity, as in the IFE.
	
	In what follows, we express all the possible polarizations of the light propagating along the $z$ axis using Jones vectors writing as:
	
	\begin{equation}\label{eq:jones}
		\bm{u} = 
		\begin{bmatrix}
			\cos{\theta} \\
			\sin{\theta} e^{i\phi} \\
			0
		\end{bmatrix}.
	\end{equation}

	Using Eq. \ref{eq:jones}, arbitrarily defined linear polarizations located in the $xy$ plane are defined by $\phi = 0$ and any value of $\theta$. On the other hand, the two circular polarizations, $\sigma^\pm$ are obtained for $\phi = \pm \frac{\pi}{2}$ and for $\theta = \frac{\pi}{4}$.

	\subsubsection{Isotropic media} \label{sec:iso}
	
	In isotropic media, the dielectric tensor writes as:
	
	\begin{equation}\label{eq:iso_dielectric}
		\tilde{\varepsilon} = 
		\left(
			\begin{matrix}
				\varepsilon'_{xx} + i\varepsilon''_{xx} & 0 & 0\\
				0 & \epsilon'_{xx} + i\varepsilon''_{xx} & 0\\
				0 & 0 & \varepsilon'_{xx} + i\varepsilon''_{xx}
			\end{matrix}
		\right)
	\end{equation}

	Just as the energy absorption of light by dissipative media (see Annex \ref{an:number_abs_photons}), the rate of absorption of orbital angular momentum, or magnetization, is directly proportional to the imaginary part of $\tilde{\varepsilon}$:
	
	\begin{equation}\label{eq:torque_iso}
		\frac{d \bm{m}_\text{abs}}{d F} = - 2 \frac{\gamma}{c} \varepsilon''_{xx} \sin{\left(\phi \right)} \sin{\left(\theta \right)} \cos{\left(\theta \right)} \bm{u}_z.
	\end{equation}

	Indeed, Eq. \ref{eq:torque_iso} shows that a torque is applied by non-linearly polarized light, \ie\ when $\phi \neq n \pi$ and $\theta \neq n \frac{\pi}{2}$.  It takes its maximum amplitude for circularly polarized light and its direction reverses with the helicity such as:
	
	\begin{equation}\label{eq:torque_iso_circ}
		\frac{d{\bm{m}_{\text{abs}, \sigma^\pm}}}{d F} = \mp \frac{\gamma}{c} \varepsilon''_{xx} \bm{u}_z.
	\end{equation}
	
	Interestingly, by dividing Eq. \ref{eq:torque_iso_circ} by the amount of absorbed photons per unit of fluence as expressed in Eq. \ref{eq:n_abs_photon}, we obtain:
	
	\begin{equation}
		\frac{d \bm{m}_{\text{abs}, \sigma^\pm}}{d n} = \mp \gamma \hbar \bm{u}_z,
	\end{equation}
	
	thus showing that, upon absorption of a circularly polarized photon, an angular momentum of $\hbar$ is indeed transferred to the electronic system.
	
\begin{figure}
		\includegraphics[scale=0.27]{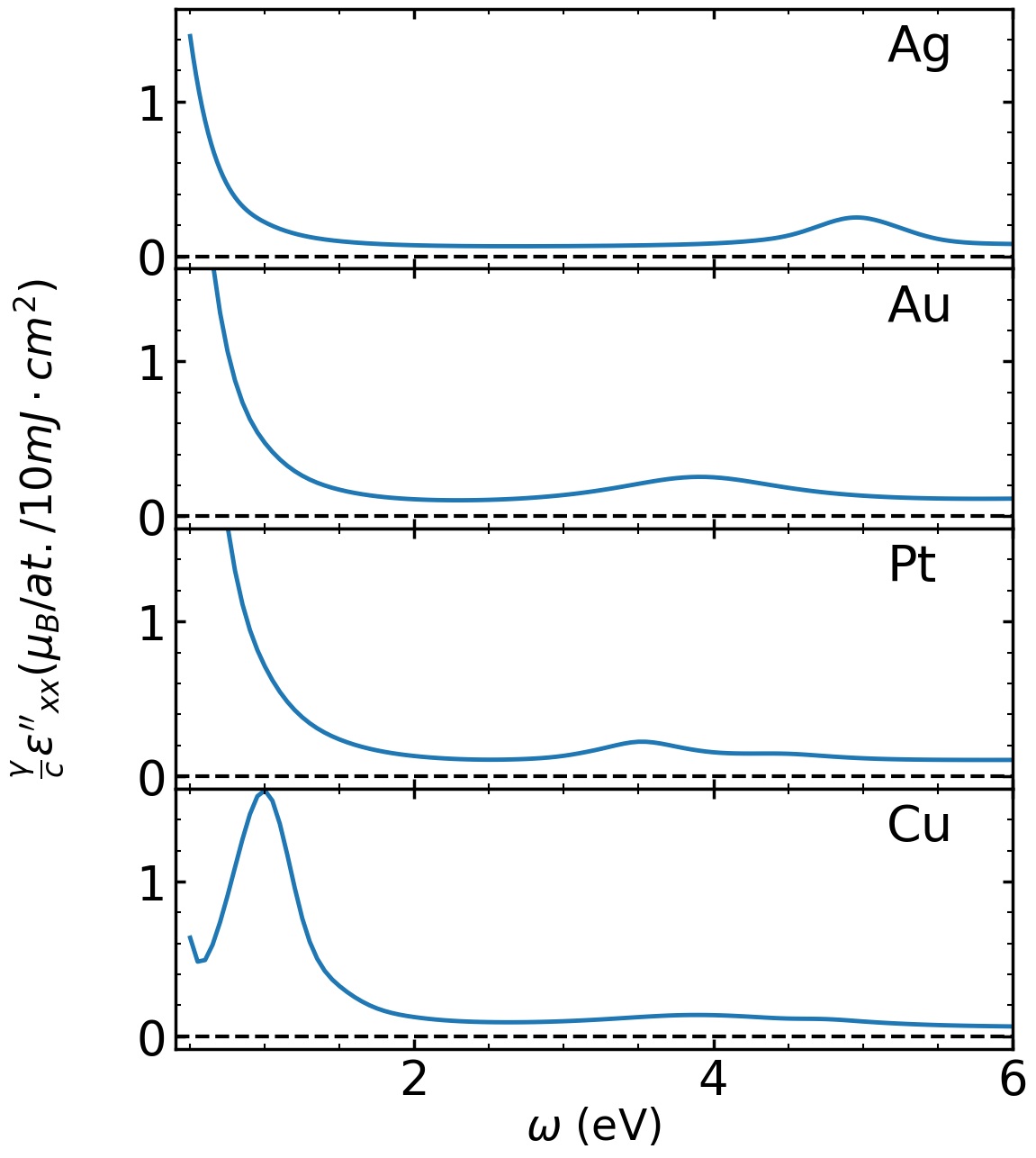}
		\includegraphics[scale=0.27]{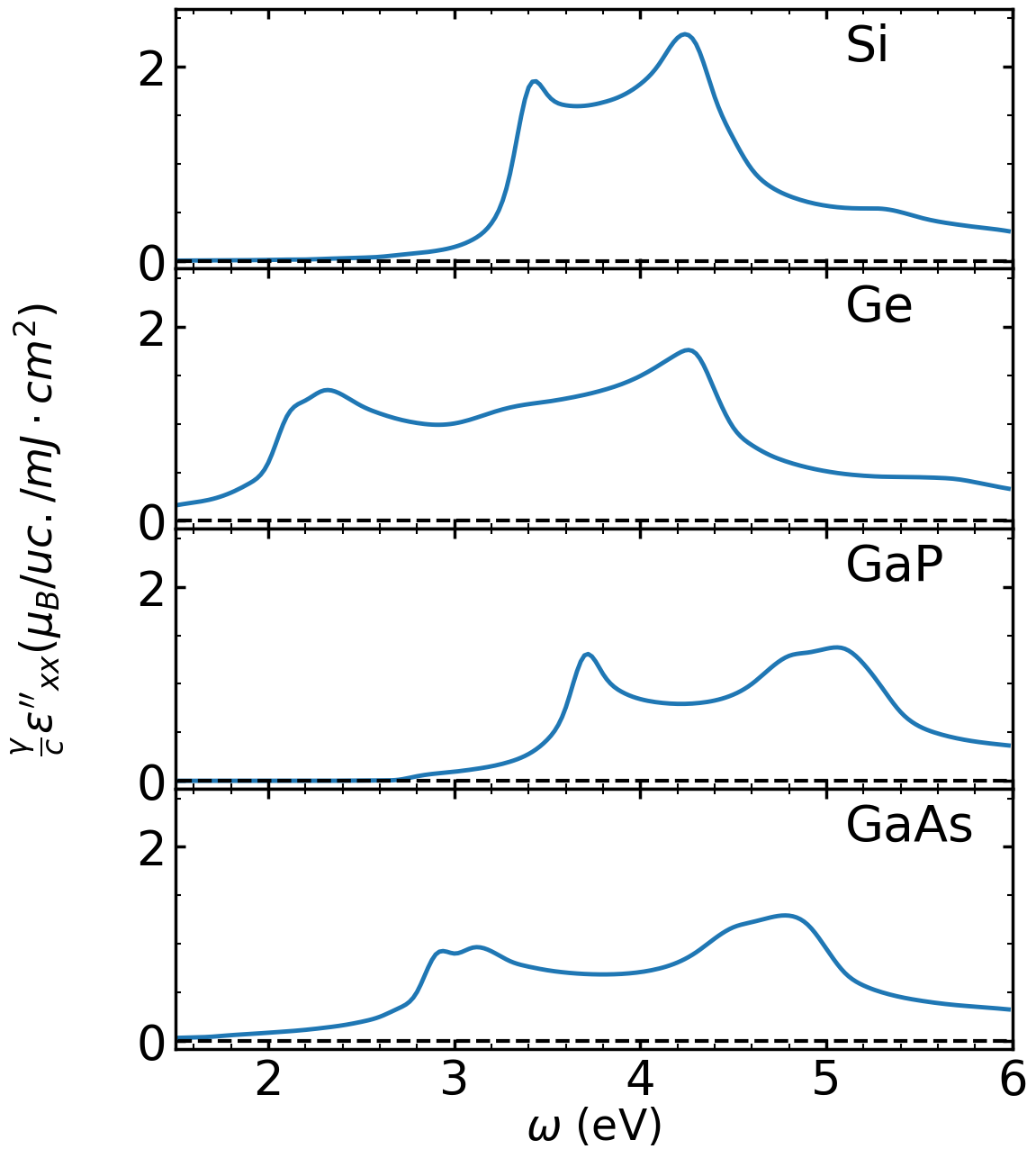}
		\caption{Top: absorbed magnetic moment in $\mu_B$ per atom and per 10mJ cm$^{-2}$ (fluence) in Ag, Au, Pt and Cu, calculated using the experimentally measured dielectric functions from Ref. \cite{werner_optical_2009}. Bottom: absorbed magnetic moment in $\mu_B$ per conventional unit cell and per mJ cm$^{-2}$ (fluence) in Si, Ge, GaP and GaAs, calculated using the experimentally measured dielectric functions from Ref. \cite{aspnes_dielectric_1983}.\label{fig:iso}}
	\end{figure}

	Also, we note that in this case, the second harmonic generation does not contribute to the induced torque, and that, as $\bm{m}_\text{abs}$ does not depend on $\varepsilon'_{xx}$, orbital angular momentum cannot be absorbed in the absence of dissipation.
	These facts are well illustrated in Fig. \ref{fig:iso}, where we used experimental measures of the dielectric function is various isotropic compounds. Indeed, metals are absorbing angular momentum at all photon energy, even though the phenomenon is particularly strong at low energy where the contribution of intraband transition (Drude like), is dominant. We note a singularity in Cu, which presents a peak of absorption at $\approx1$eV. On the other hand, in semi-conductor, the absorption of angular momentum takes place only at energies greater than the gap, where absorption is possible.

	\subsubsection{Uniaxial media}
	
	In uniaxial media, the dielectric tensor writes as:
	
	\begin{equation}
		\tilde{\varepsilon}(\Omega) = 
		\left(
		\begin{matrix}
			\tilde{\varepsilon}'_{xx} + i\tilde{\varepsilon}''_{xx} & 0 & 0\\
			0 & \tilde{\varepsilon}'_{yy} + i\tilde{\varepsilon}''_{yy} & 0\\
			0 & 0 & \tilde{\varepsilon}'_{yy} + i\tilde{\varepsilon}''_{yy}
		\end{matrix}
		\right),
	\end{equation}

	thus leading to the following absorbed density of angular momentum per unit of fluence:
	
	\begin{equation}\label{eq:uniax_gen}
		\begin{split}
			\frac{d{\bm{m}_{\text{abs}}}}{d F} =  - \frac{\gamma}{c} \left(\varepsilon'_{xx} - \varepsilon'_{yy}\right) \sin{\left(\theta \right)} \cos{\left(\theta \right)} \Big[ \cos{\left(\phi \right)}\\
			+ \cos{\left(\phi\right)} \cos{\left(2 \Omega t\right)} - \sin{\left(\phi\right)} 	\sin{\left(2 \Omega t\right)} \Big] \bm{u}_z\\
			- \frac{\gamma}{c} \left(\varepsilon''_{xx} + 	\varepsilon''_{yy}\right) \sin{\left(\theta \right)} \cos{\left(\theta \right)} \sin{\left(\phi\right)} \bm{u}_z \\
			+ \frac{\gamma}{c} \left(\varepsilon''_{xx} - 	\varepsilon''_{yy}\right) \sin{\left(\theta \right)} \cos{\left(\theta \right)} \Big[ \sin{\left(\phi\right)} \cos{\left(2 \Omega t\right)}\\
			+ \cos{\left(\phi\right)} \sin{\left(2 \Omega t\right)} \Big] \bm{u}_z
		\end{split}
	\end{equation}
	
	As shown by the first two lines of Eq. \ref{eq:uniax_gen}, and contrary to the isotropic case, a component proportional to $\varepsilon'_{xx} - \varepsilon'_{yy}$, which are real parts of the dielectric tensor is now present. Consequently, in such media, orbital angular momentum can be absorbed from light, even in absence of energy absorption. Contrary to the isotropic case, this phenomenon is proportional to $\cos{\left(\phi\right)}$, \ie\ its amplitude is maximal for linearly polarized light ($\phi = n \pi$) sitting in the directions $\theta = \pm\frac{\pi}{4}$. Moreover, the sign of this effect reverses with $\theta$. In such a polarization a second harmonic is also generated such that the total torque writes as: 
	
	\begin{equation}\label{eq:uni_non_dis_lin}
		\frac{d \bm{m}_{\text{abs}, \pi^{\theta = \pm\frac{\pi}{4}}}^{\varepsilon'_{xx} - \varepsilon'_{yy}}}{d F} = \mp \frac{\gamma}{2 c} \left( \varepsilon'_{xx} - \varepsilon'_{yy} \right) \left(1 + \cos{\left(2 \Omega t\right)}\right) \bm{u}_z
	\end{equation}
	
	This phenomenon cancels for $\theta = n \frac{\pi}{2}$, \ie\ whenever the polarization of the light is linear and along $x$ or $y$, corresponding to the eigenstates of the dielectric tensor.
	
	For circular polarizations, the real part of the dielectric tensor is only responsible for a second harmonic generation, which phase factor depends on the helicity:
	
	\begin{equation}\label{eq:uni_non_dis_circ}
		\frac{d \bm{m}_{\text{abs}, \sigma^{\pm}}^{\varepsilon'_{xx} - \varepsilon'_{yy}}}{d F} = \pm\frac{\gamma}{2 c} \left( \varepsilon'_{xx} - \varepsilon'_{yy} \right) \sin \left(2 \Omega t\right) \bm{u}_z
	\end{equation}

	On top of this non-dissipative contribution, the presence of dissipation brings a static torque which has the same characteristics as in the isotropic absorptive media, \ie\ cancels for linearly polarized light, has its maximal amplitude for circular polarizations, and reverses its sign with the latter, such as:
	
	\begin{equation}\label{eq:uni_disp_circ}
		\frac{d \bm{m}_{\text{abs}, \sigma^{\pm}}^{\varepsilon''_{xx} + \varepsilon''_{yy}}}{d F} = \mp \frac{\gamma}{2 c} \left( \varepsilon''_{xx} + \varepsilon''_{yy} \right) \bm{u}_z
	\end{equation}

	 Moreover, and as shown by the last two lines of Eq. \ref{eq:uniax_gen}, the complex part of the dielectric tensor also induces a second harmonic proportional to $\varepsilon''_{xx} - \varepsilon''_{yy}$. As the second harmonic proportional to $\varepsilon'_{xx} - \varepsilon'_{yy}$, the latter cancels for linearly polarized light within the $x$ and $y$ directions and takes its maximum amplitude for circularly polarized light, expressing as:
	 
	 \begin{equation}\label{eq:uni_disp_sec_harm_circ}
	 	\frac{d \bm{m}_{\text{abs}, \sigma^{\pm}}^{\varepsilon''_{xx} - \varepsilon''_{yy}}}{d F} = \pm\frac{\gamma}{2 c} \left( \varepsilon''_{xx} - \varepsilon''_{yy} \right) \sin(2 \Omega t)\bm{u}_z,
	 \end{equation}
	 
	 as well as for linearly polarized light when $\theta= \pm\frac{\pi}{4}$, such as:
	 
	 \begin{equation}\label{eq:uni_disp_sec_harm_lin}
	 	\frac{d \bm{m}_{\text{abs}, \pi^{\theta = \pm\frac{\pi}{4}}}^{\varepsilon''_{xx} - \varepsilon''_{yy}}}{d F} = \mp \frac{\gamma}{2 c} \left( \varepsilon''_{xx} - \varepsilon''_{yy} \right) \cos(2 \Omega t) \bm{u}_z
	 \end{equation}

	\begin{figure}
		\includegraphics[scale=0.27]{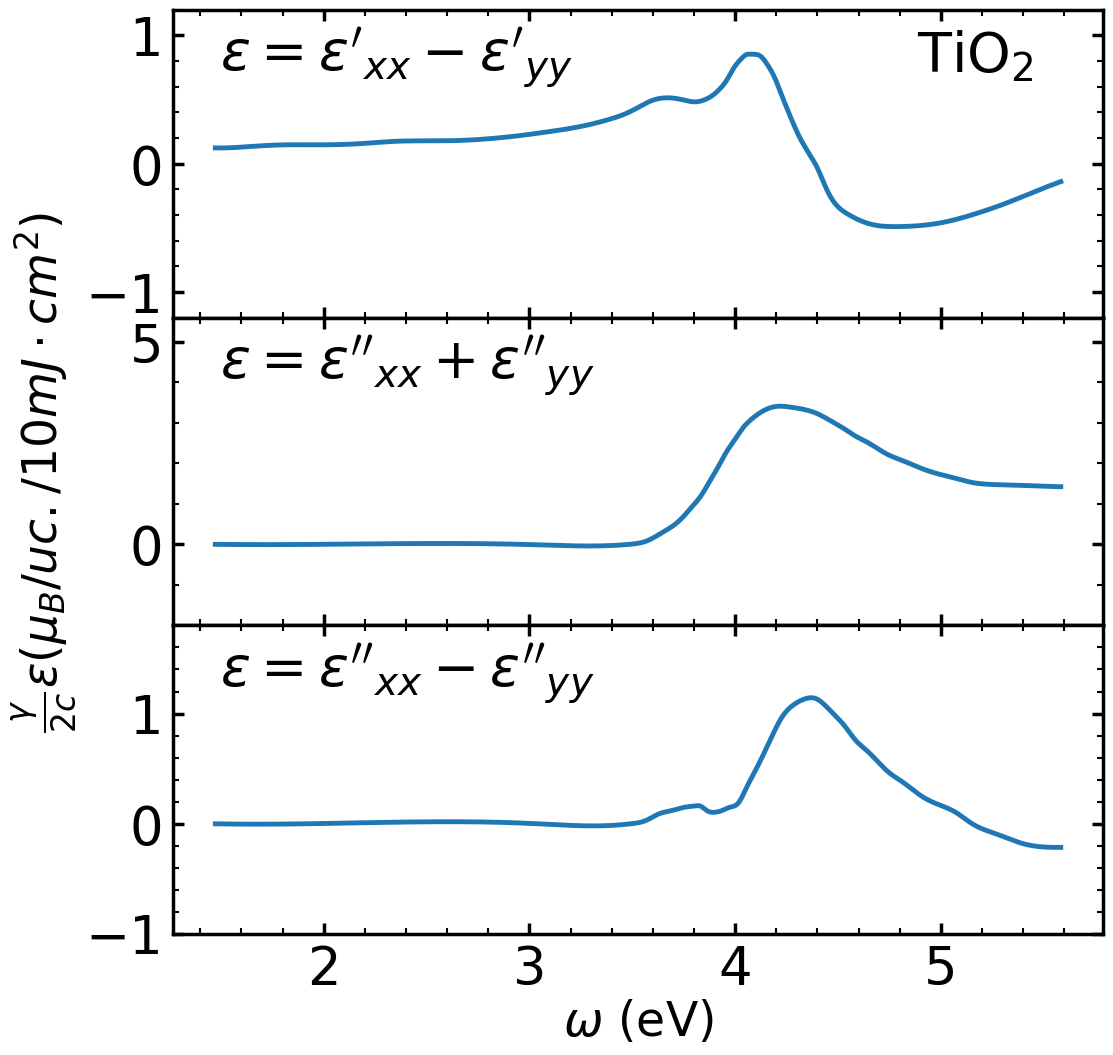}
		\includegraphics[scale=0.27]{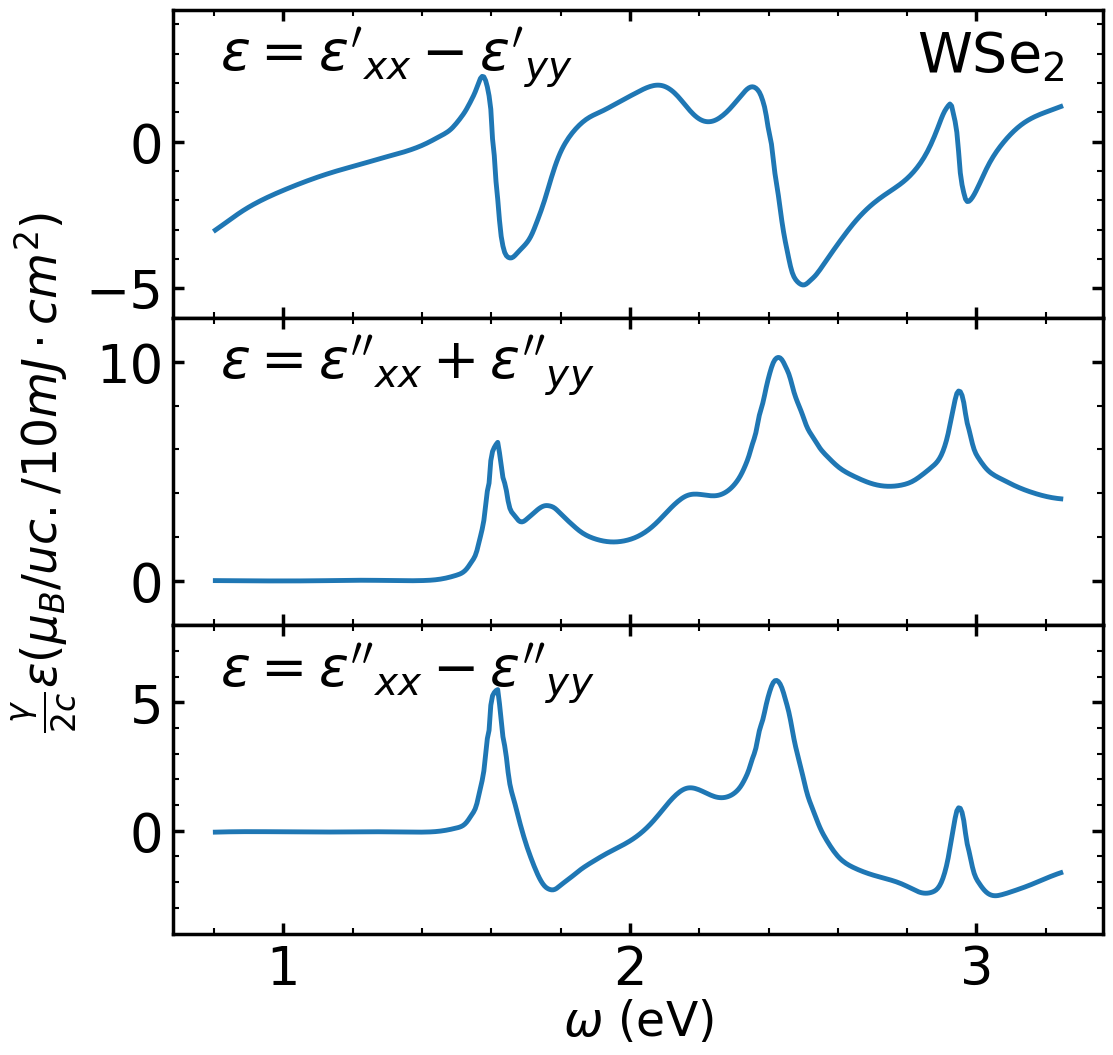}
		\caption{Amplitude of the contributions to the absorbed orbital magnetic moment per unit cell of rutile TiO$_2$ using experimental dielectric functions measured by Jellison \ea\cite{jellison_spectroscopic_2003}, and per unit cell of monolayer WSe$_2$ using experimental dielectric functions measured by Chu \ea\cite{chu_birefringent_2020}.\label{fig:uniaxe}}
	\end{figure}

	Using experimental measurements of the dielectric function in rutile TiO$_2$\cite{jellison_spectroscopic_2003}, which space group is P4$_2$/mnm and holds 12 atoms within its conventional cell, and monolayer WSe$_2$\cite{chu_birefringent_2020} which space group is P6$_3$/mmc and holds 6 atoms within its unit cell, we compute the amplitude of the different aforementioned phenomena, as seen in Fig. \ref{fig:uniaxe}. As both of these compounds present a gap, the dissipative part, proportional to $\varepsilon''_{xx} + \varepsilon''_{yy}$ is not contributing up to $\approx 3.5$eV in TiO$_2$, and $\approx 1.5$eV in WSe$_2$. On the other hand, even below these energies, the non-dissipative contributions, proportional to $\varepsilon'_{xx} - \varepsilon'_{yy}$ are allowing for the absorption of orbital angular momentum from the light within the aforementioned conditions of Eq. \ref{eq:uni_non_dis_lin} and Eq. \ref{eq:uni_non_dis_circ}. The sign of this component, which can lead to the absorption of orbital angular momentum even using a linearly polarized light, depends on the energy of the light in both compounds. In the two compounds, when above the gap energy, the largest contribution comes from the regular, static part of the torque, proportional to $\varepsilon''_{xx} + \varepsilon''_{yy}$ represented by Eq. \ref{eq:uni_disp_circ}, and which is maximal for circular polarizations. We notice however that, even above the absorption threshold, the magnetization induced by the static part of the torque, proportional to $\varepsilon'_{xx} - \varepsilon'_{yy}$, cannot be neglected in front of its dissipative counterpart. Finally, the same can be said regarding the second harmonics generations, which are proportional to $\varepsilon''_{xx} + \varepsilon''_{yy}$, and thus only exist above the absorption threshold. 

	\subsubsection{Monoclinic media}
	
	In monoclinic media, the dielectric tensor writes as:
	
	\begin{equation}
		\tilde{\varepsilon}(\Omega) = 
		\left(
		\begin{matrix}
			\tilde{\varepsilon}'_{xx} + i \tilde{\varepsilon}''_{xx} & \tilde{\varepsilon}'_{xy} + i \tilde{\varepsilon}''_{xy} & 0\\
			\tilde{\varepsilon}'_{xy} + i \tilde{\varepsilon}''_{xy} & \tilde{\varepsilon}'_{xx} + i \tilde{\varepsilon}''_{xx} & 0\\
			0 & 0 & \tilde{\varepsilon}'_{xx} + i \tilde{\varepsilon}''_{xx}
		\end{matrix}
		\right)
	\end{equation}
	
	Such that the induced torque is:
	
	\begin{equation}\label{eq:torque_abs_anis}
		\begin{split}
			\frac{d \bm{m}_\text{abs}}{d F}(t) =  - 2\frac{\gamma}{c} \varepsilon''_{xx} \sin{\left(\phi \right)} \sin{\left(\theta \right)} \cos{\left(\theta \right)} \bm{u}_z\\
			- \frac{\gamma}{c} \varepsilon'_{xy} \big[ \cos{\left(2 \theta \right)} + \left(\sin^2 \theta \cos 2 \phi - \cos^2 \theta\right) \cos 2\Omega t\\
			- 2 \sin^2 \theta \sin \phi \cos \phi \sin 2 \Omega t\big] \bm{u}_z\\
			- \frac{\gamma}{c} \varepsilon''_{xy} \big[\left(\sin^2 \theta \cos 2 \phi - \cos^2 \theta\right) \sin 2\Omega t\\
			- 2 \sin^2 \theta \sin \phi \cos \phi \cos 2 \Omega t\big] \bm{u}_z.
		\end{split}
	\end{equation}
	
	The first line of Eq. \ref{eq:torque_abs_anis} is the component of the torque induced by the dissipative part of the diagonal elements, and is similar to the aforementioned isotropic case, given by Eq. \ref{eq:torque_iso}.
	In addition, two components proportional to the non-diagonal elements, $\tilde{\varepsilon}'_{xy}$ and $\tilde{\varepsilon}''_{xy}$, due to the monoclinic nature of the compound, are appearing. Remarkably, only the former brings a static contribution to the torque, and therefore leads to the absorption of angular momentum. It does not depend on the ellipticity of the light, \ie\ on $\phi$, and cancels when $\theta = \frac{\pi}{4} + n\frac{\pi}{2}$, which when linearly polarized, correspond to the eigenstate polarization of the dielectric tensor. On the other hand, its amplitude is maximal along the directions $\theta = 0$, and $\theta = \pi$, \ie\ in the case of a linearly polarized light along the $x$ and $y$ direction, and writes as:
	
	\begin{equation}
		\frac{d \bm{m}_{\text{abs}, \pi=\left\{0,\pi\right\}}^{\text{stat.}}}{d F} = \mp \frac{\gamma}{c} \varepsilon'_{xy} \bm{u}_z.
	\end{equation}
	
 	Still for a linearly polarized light, the second harmonic generation proportional to $\tilde{\varepsilon}'_{xy}$ has the same properties as the aforementioned static component, thus expressing as:
	
	\begin{equation}
		\frac{d\bm{m}_{\text{abs}, \pi=\left\{0,\pi\right\}}^{2^\text{nd} \text{h.}}}{d F} (t) = \mp \frac{\gamma}{c} \varepsilon'_{xy} \cos \left(2 \Omega t\right) \bm{u}_z
	\end{equation}

	Thus, when the polarization is eigenvalue of the dielectric tensor, the whole torque cancels. On the other hand, when circularly polarized, the second harmonic generation writes as:
	
	\begin{equation}
		\frac{d \bm{m}_{\text{abs}, \sigma^\pm}^{2^\text{nd} \text{h.}}}{d F}(t) = - \frac{\gamma}{c} \varepsilon'_{xy} \cos \left(2 \Omega t\right) \bm{u}_z,
	\end{equation}

	\ie\ is the same for both helicities of the light.
	
	Finally, the second harmonics proportional to $\tilde{\varepsilon}''_{xy}$ present exactly the same properties as the ones proportional to its non-dissipative counterpart, with the addition of a de-phasing of $-\frac{\pi}{2}$, \ie:
	
	\begin{equation}
		\frac{d \bm{m}_{\text{abs}, \pi=\left\{0,\pi\right\}}^{2^\text{nd} \text{h.}}}{d F} (t) = \mp \frac{\gamma}{c} \varepsilon''_{xy} \sin \left(2 \Omega t\right) \bm{u}_z,
	\end{equation}

	and
	
	\begin{equation}
		\frac{d\bm{m}_{\text{abs}, \sigma^\pm}^{2^\text{nd} \text{h.}}}{d F} (t) = - \frac{\gamma}{c} \varepsilon''_{xy} \sin \left(2 \Omega t\right) \bm{u}_z.
	\end{equation}
	
	\begin{figure}
	\includegraphics[scale=0.27]{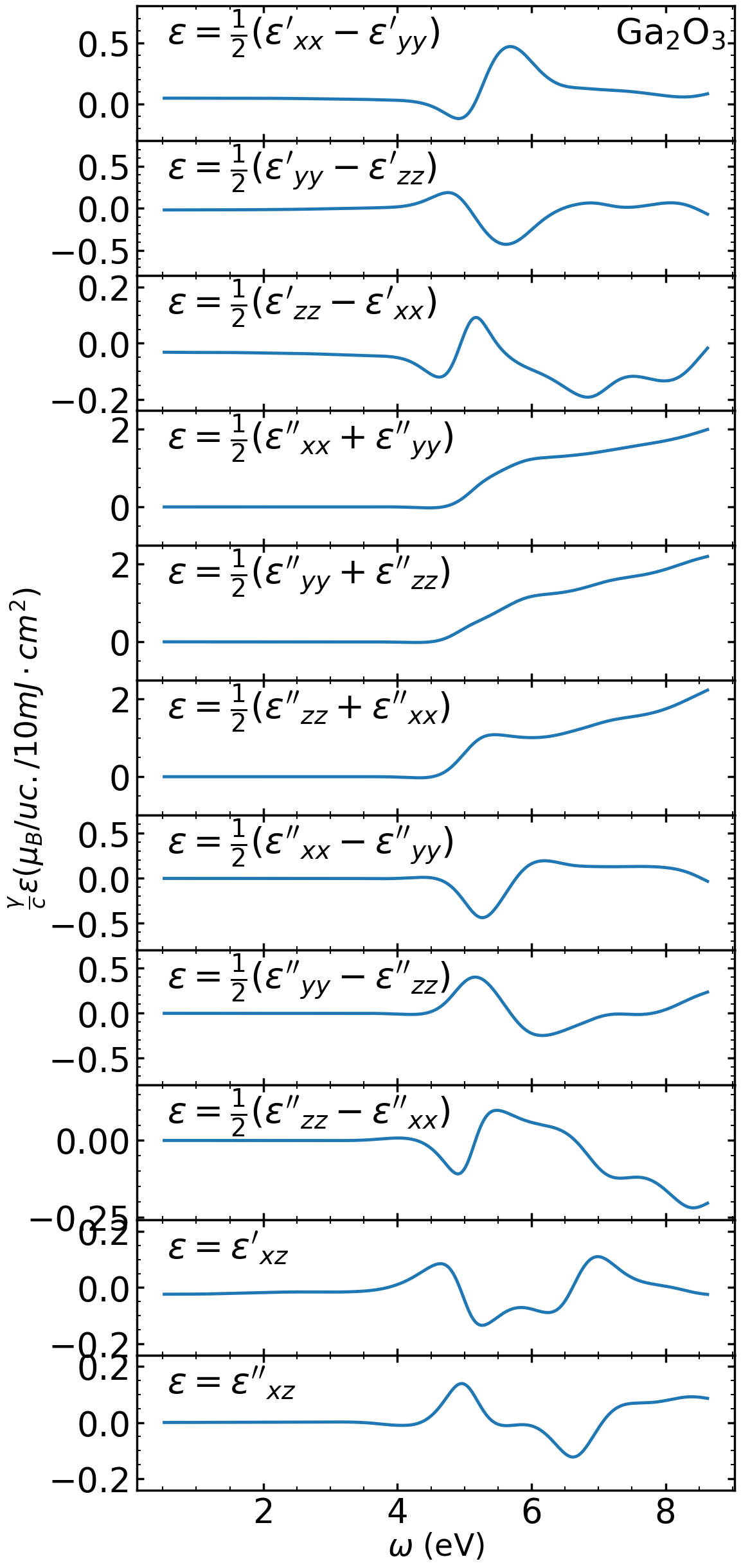}
	\caption{Amplitude of the contributions to the absorbed orbital magnetic moment per unit cell of Ga$_2$O$_3$ using experimental dielectric functions measured by Sturm \ea\cite{sturm_dielectric_2015}.\label{fig:Ga2O3}}
	\end{figure}

	To illustrate this case we use the work of Sturm \ea\cite{sturm_dielectric_2015}, who experimentally measured the dielectric tensor of the monoclinic ($\beta$) phase of Ga$_2$O$_3$. It possesses 20 atoms per unit cell, which volume represents 211.47 A$^3$. Using the same orientation as Sturm \ea, its dielectric tensor writes as:
	
	\begin{equation}
		\tilde{\varepsilon}(\Omega) = 
		\left(
		\begin{matrix}
			\tilde{\varepsilon}_{xx} & 0 & \tilde{\varepsilon}_{xz}\\
			0 & \tilde{\varepsilon}_{yy} & 0\\
			\tilde{\varepsilon}_{xz} & 0 & \tilde{\varepsilon}_{zz}
		\end{matrix}
		\right)
	\end{equation}

	Consequently, in addition of the non-diagonal term, $\tilde{\varepsilon}_{xz}$, and depending on the orientation of the perturbation, \ie\ of the plane in which the electric field is evolving through time ($xy$, $yz$ or $zx$), the phenomena represented by Eq. \ref{eq:uni_non_dis_lin}, \ref{eq:uni_non_dis_circ}, \ref{eq:uni_disp_circ}, \ref{eq:uni_disp_sec_harm_circ}, and \ref{eq:uni_disp_sec_harm_lin} will triggered at three different amplitudes, all of which are represented in the first 9 graphs of Fig. \ref{fig:Ga2O3}. As, in general, the values taken by the dielectric function of Ga$_2$O$_3$ are small as compared to the other compounds studied \textit{supra}, the amplitude found for the different aforementioned phenomena are equally smaller. However, and as usual, the absorption of angular momentum proportional to the dissipative diagonal terms (here, their sums), occurring in the case of a circularly polarized light are the largest. Moreover, in this case, their value noticeably depends on the plane containing the circularly polarized electric field. The latter dependence is only exacerbated in the case of phenomena proportional to the difference of the diagonal elements. Finally, and most directly related to the case of monoclinic media, the phenomena proportional to $\varepsilon'_{xz}$ and $\varepsilon''_{xz}$ have the same order of magnitude, and therefore cannot be neglected with respect to the ones proportional to differences of diagonal elements.

	\subsubsection{Isotropic magnetic material} \label{sec:iso_mag}
	
	Lastly, we treat the case of an isotropic magnetic material. Its dielectric tensor writes as:
	
	\begin{equation}
		\tilde{\varepsilon}(\Omega) = 
		\left(
		\begin{matrix}
			\tilde{\varepsilon}'_{xx} + i\varepsilon''_{xx} & i (\tilde{\varepsilon}'_{xy} + i \tilde{\varepsilon}''_{xy}) & 0\\
			-i (\tilde{\varepsilon}'_{xy} + i \tilde{\varepsilon}''_{xy}) & \tilde{\varepsilon}'_{xx} + i\tilde{\varepsilon}''_{xx} & 0\\
			0 & 0 & \tilde{\varepsilon}'_{xx} + i\tilde{\varepsilon}''_{xx}
		\end{matrix}
		\right)
	\end{equation}

	\begin{equation}\label{eq:torque_abs_mag}
		\begin{split}
			\frac{d \bm{m}_\text{abs}}{d F}(t) &= - 2\frac{\gamma}{c} \varepsilon''_{xx} \sin{\left(\phi \right)} \sin{\left(\theta \right)} \cos{\left(\theta \right)}\bm{u}_z\\
			& + \frac{\gamma}{c}\varepsilon'_{xy} \big[\left(\cos^2 \left(\theta\right) + \cos \left( 2 \phi \right) \sin^2 \left(\theta\right)\right) \sin \left(2 \Omega t\right)\\
			& + 2 \sin \left(\phi\right) \cos\left(\phi\right) \sin^2 \left(\theta\right) \cos\left(2 \Omega t\right)
			\big] \bm{u}_z\\
			& + \frac{\gamma}{c} \varepsilon''_{xy} \bm{u}_z \\
			& + \frac{\gamma}{c} \varepsilon''_{xy} \big[ -\left( \cos^2(\theta) + \cos(2\phi) \sin^2(\theta)\right) \cos\left(2 \Omega t\right)\\
			& + 2\sin^2\left(\theta\right)\cos\left(\phi\right)\sin\left(\phi\right)\sin\left(2\Omega t\right)\big] \bm{u}_z 
		\end{split}
	\end{equation}

	Here, and as in the previous cases, in a non-dissipative medium and when the light is polarized in an eigenvalue of the dielectric tensor, which here are the two helicities of the light, no torque is induced and we have:
	
	\begin{equation}\label{eq:}
		\frac{d \bm{m}_\text{abs}^{\text{non-dis.}}}{d F} = \bm{0}
	\end{equation}
	
	On the other hand, when the light is not in such a state, and as shown by the term proportional to $\varepsilon'_{xy}$, second harmonics will be generated. When linearly polarized, the latter writes as:
	
	\begin{equation}\label{eq:real_IFE}
		\frac{d \bm{m}_\text{abs}^{\varepsilon'_{xy}}}{d F}(t) = \frac{\gamma}{c} \varepsilon'_{xy} \sin(2\Omega t)\bm{u}_z.
	\end{equation}

	Here we note that it is precisely the configuration in which the Faraday effect takes place, \ie\ a linearly polarized light, which axis of polarization rotates as it propagates along the direction of the magnetization. According to the Onsager reciprocal relations, it would therefore appear that the IFE is the generation of a time-dependent magnetic moment by such a linearly polarized light, as shown by Eq. \ref{eq:real_IFE}, rather than the generation of a magnetization density by circularly polarized light as originally pinned by Pitaevskii \cite{pitaevskii_electric_1961}. Indeed, this is readily hinted by the fact that $\varepsilon'_{xy}$, which is precisely the component of the dielectric tensor that dictates the magnitude of rotation of the light in the Faraday effect occurring in non-lossy materials, is also the component to which the second harmonic generation described by Eq. \ref{eq:real_IFE} is proportional to.\\
	The presence of dissipation brings many additional contributions. The first one, represented by the first line of Eq. \ref{eq:torque_abs_mag} is the component of the torque induced by the dissipative part of the diagonal elements, and is similar to the aforementioned isotropic case, given by Eq. \ref{eq:torque_iso}.
	The second one, also static, writing as:
	
	\begin{equation}
		\frac{d \bm{m}_\text{abs}^{\varepsilon''_{xy}, \text{stat.}}}{d F} = \frac{\gamma}{c} \varepsilon''_{xy} \bm{u}_z,
	\end{equation}
	
	is the same for any polarization of the light and is due to both the presence of a magnetic moment and of dissipation. It shows that, in such materials, even linearly polarized light can induce the absorption of a magnetic moment by the matter.
	Also proportional to $\varepsilon''_{xy}$, and as shown by the last two lines of Eq. \ref{eq:torque_abs_mag}, the dissipative, out of diagonal part of the response also participates to the generation of second harmonics which, in the case of a linear polarization writes as:
	
	\begin{equation}
		\frac{d \bm{m}_\text{abs}^{\varepsilon''_{xy}, \text{sec. h.}}}{d F}(t) = \frac{\gamma}{c} \varepsilon''_{xy} \cos(2\Omega t)\bm{u}_z.
	\end{equation}

	As expected because of rotational symmetry around the $z$ axis, and contrary to the case of a monoclinic material, the induced torque does not depend on $\theta$.
	
\begin{figure}
		\includegraphics[scale=0.26]{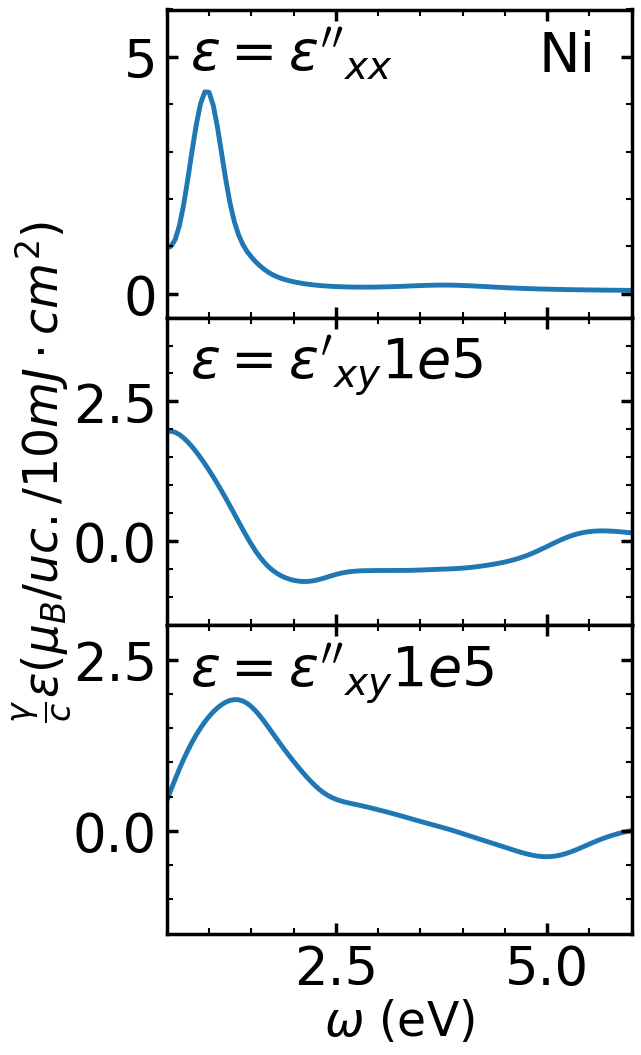}\includegraphics[scale=0.26]{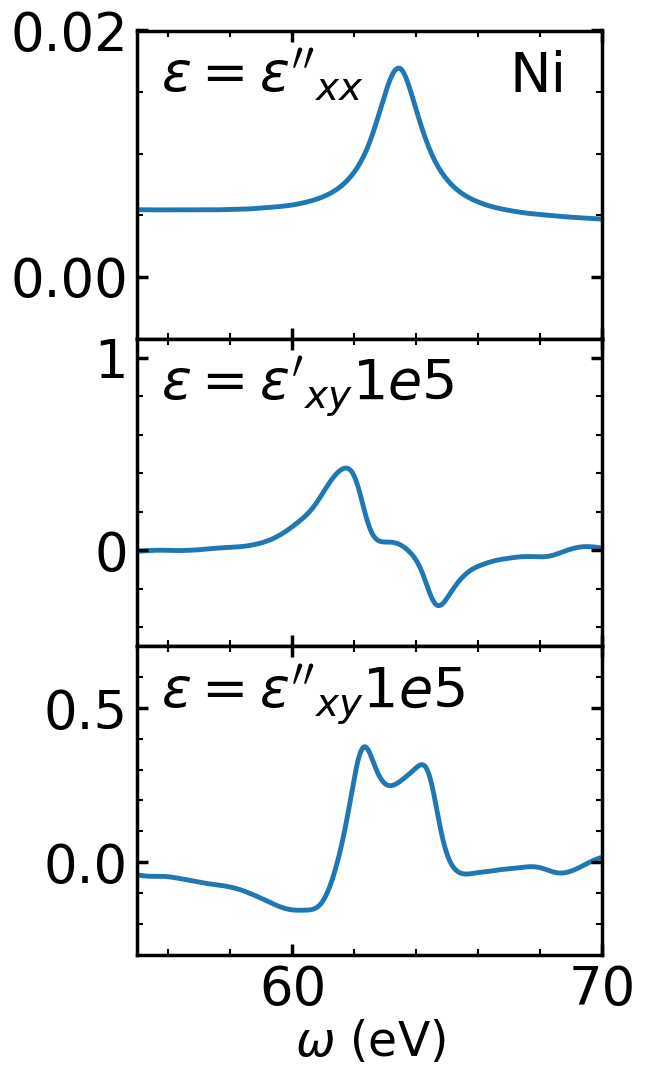}\\
		\includegraphics[scale=0.26]{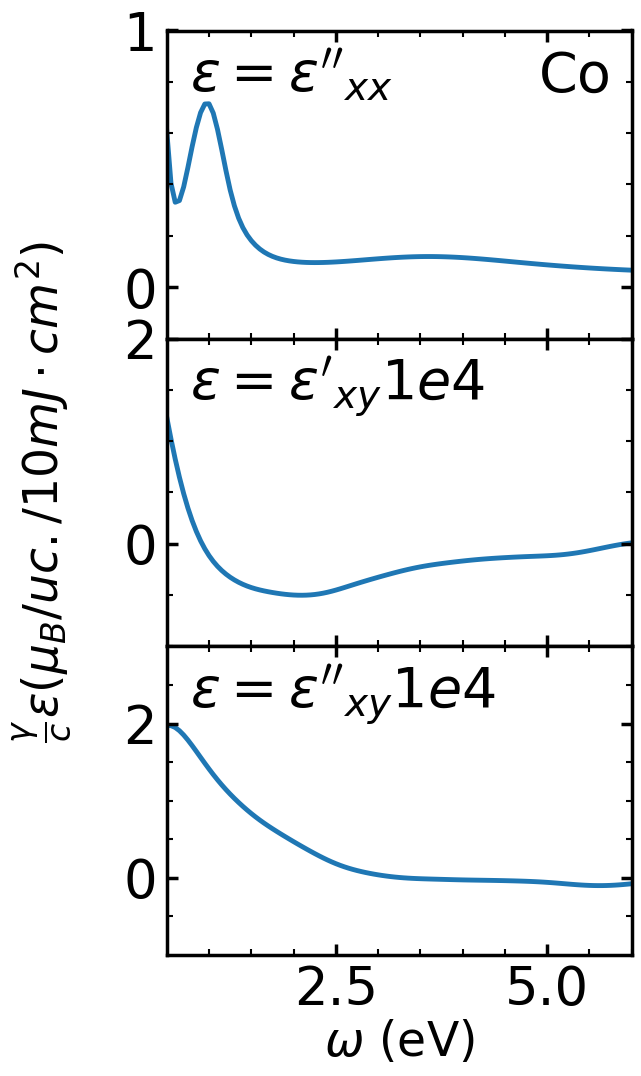}\includegraphics[scale=0.26]{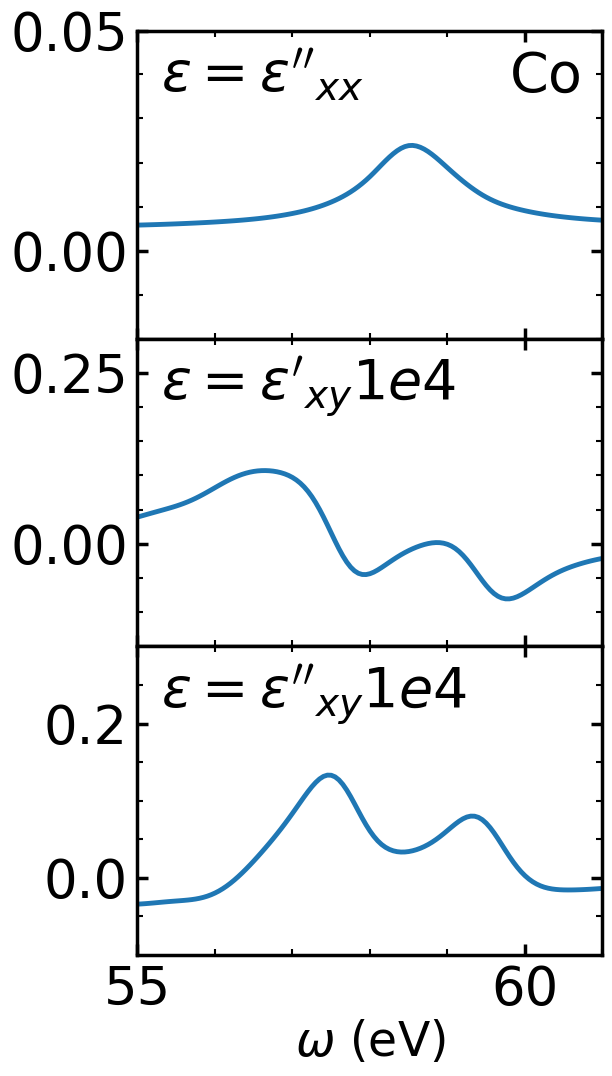}\\
		\includegraphics[scale=0.26]{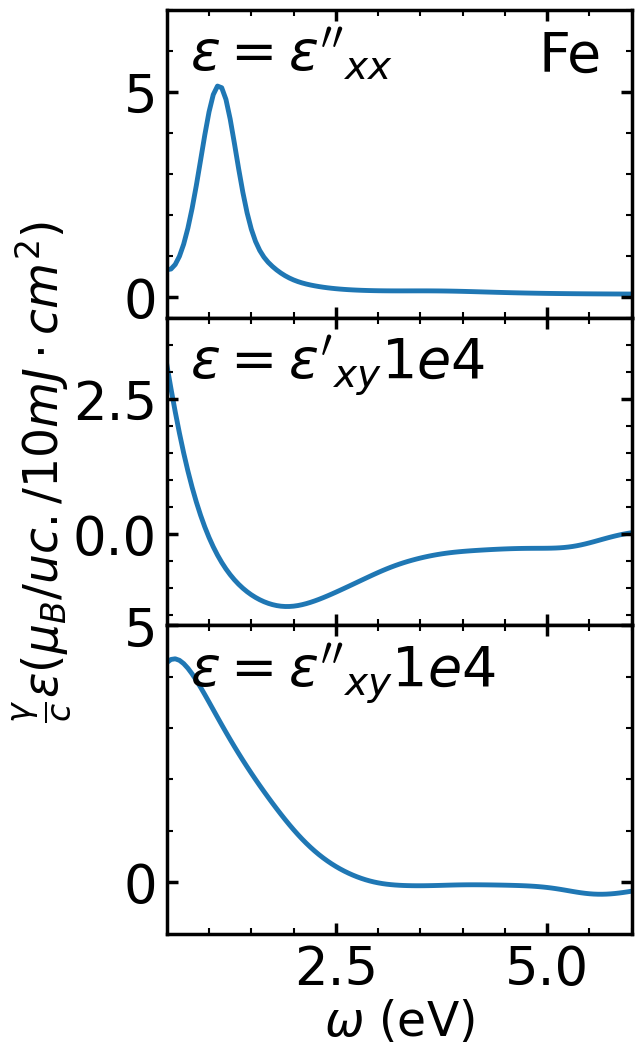}\includegraphics[scale=0.26]{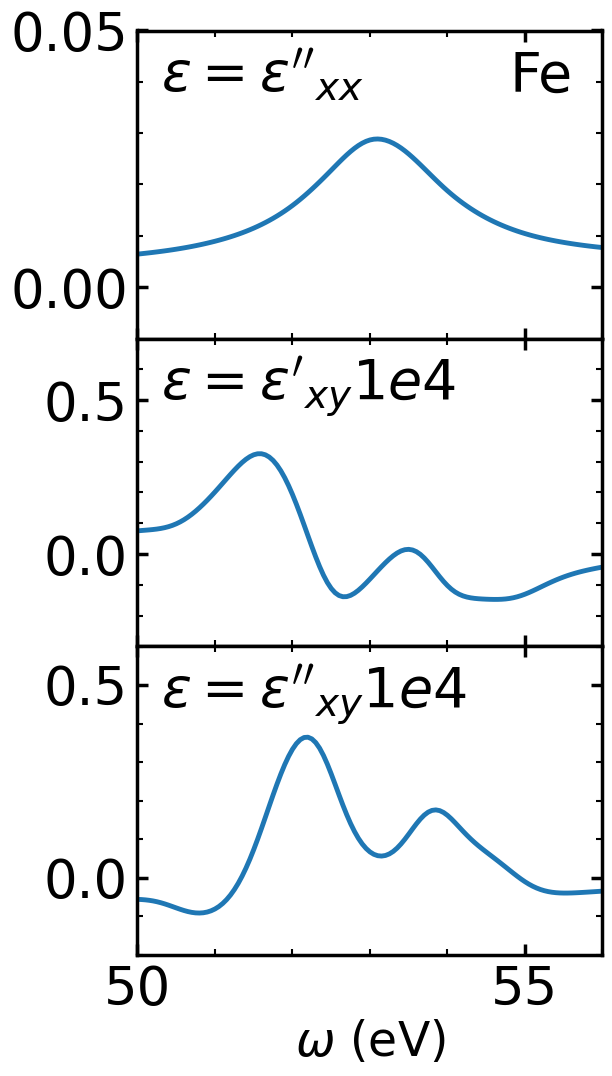}
		\caption{Amplitude of the contributions to the absorbed orbital magnetic moment per unit cell in the elemental ferromagnets Ni, Co and Fe using experimental dielectric functions measured by Werner \ea\cite{werner_optical_2009} \ea\cite{sturm_dielectric_2015}.\label{fig:mag}}
	\end{figure}

	Fig. \ref{fig:mag} presents the amplitude of all the aforementioned phenomena in Ni, Co and Fe. While $\varepsilon''_{xx}$ is extracted from experiments\cite{werner_optical_2009}, the out-of-diagonal components are obtained from \ai\ density functional linear response calculations (see annexes). In all the compounds, and by 4 to 5 orders of magnitude, the component of the torque induced by the dissipative part of the diagonal elements dominates. Therefore, and contrary to the case of monoclinic compounds, the out-of-diagonal elements induced by the presence of a magnetic moment are negligible. This conclusion holds true as much in the optical range than at the M-edge.

	\section{Discussions}
	
	\subsection{Symmetry breaking and light-induced torque}
	
	As evidenced throughout Sec. \ref{sec:application} the light-induced torque in permitted by any symmetry-breaking arrangement of the matter, be it spatial and thus due to a deviation from the isotropic configuration, or related to the breaking of time-reversal symmetry, manifesting through the presence of dissipation and/or magnetism. Indeed, in such type of media, the polarization of light can deviate from the eigenstates of the dielectric tensor such that a static and a second harmonic torque generation can emerge.\\
	As a consequence, and differently from the light-induced angular momentum as within the IFE defined by Pitaevskii, no torque is applied by light in isotropic, transparent and non-magnetic materials. On the other hand, the largest contribution to the light-induced torque generally directly arises in lossy materials from the absorption of circularly polarized light. Interestingly, as shown in Sec. \ref{sec:iso}, and directly supporting the fact that the spin-angular momentum of light is absorbed by the electronic system, in isotropic materials the angular momentum absorbed as a consequence the integrated torque produced per absorbed circularly-polarized photon is equal to $\hbar$.\\
	On the other hand, angular momentum can also be absorbed without the presence of dissipation in uniaxial and monoclinic media. In other words, a static contribution in the induced torque can exist without a static contribution in the light-induced work. This is notably remanent of the seminal work of Beth\cite{beth_mechanical_1936} showing that, when light propagates through a birefringent material and its polarization evolves, it induces a torque that could be experimentally measured. Conversely, in non-dissipative magnetic materials, time-reversal symmetry breaking does not induce a static torque, but rather generates a dynamic, time-dependent torque that oscillates sinusoidally at the second harmonic of the incident light frequency.
	
	\subsection{Differences between the inverse Faraday effect and the light-induced torque}
	
	While discovered more than half a century ago, the inverse Faraday effect as described by Pitaevskii\cite{pitaevskii_electric_1961} has gained much interest in the past decades. Indeed, as the latter originally described the induction of a magnetization in non-magnetic transparent media proportional to the intensity of the light, and which direction depends on its helicity, it stood as a prime candidate in the explanation of a plethora of light-induced helicity-dependent effects such as the all-optical helicity-dependent switching\cite{mangin_engineered_2014,lambert_all-optical_2014}, the helicity-dependent demagnetization\cite{tsema_helicity_2016, quessab_resolving_2019} and the helicity-dependent domain wall motion\cite{quessab_helicity-dependent_2018}. 
	While the effect discussed in this work, namely the absorption of orbital angular momentum, also results in the induction of a magnetization by the light, it differs from the IFE. This is most obviously evidenced by the fact that, contrary to the latter, the absorption of angular momentum cancels in isotropic non-dissipative media. Furthermore, its magnitude is not proportional to the intensity of light, as is the IFE, but rather to its fluence. This difference, and as pinpointed in Scheid \ea\cite{scheid_light_2022} for the spin counterpart\cite{scheid_ab_2019} of the currently discussed effect, originates from the fact that, rather than being a light-induced moment as is the IFE, it is the consequence of a light-induced torque. Consequently, and just as the absorption of energy from light, the angular momentum absorbed as a result of the integrated light-induced torque stays, even after the light is gone. As such, this effect, as well as its spin counterpart\cite{scheid_ab_2019} have a lasting impact on the magnetization, that perdure long after the light is gone. Given their irreversible nature, the light-induced torques are therefore prime candidates in the explanation of the aforementioned helicity-dependent effects.
	In the light of this work one could thus re-qualify theoretical results mis-attributed to the IFE\cite{hurst_magnetic_2018, sinha-roy_driving_2020} as being the result of the light-induced torque.
	
	\subsection{Towards a new definition of the inverse Faraday effect?}
	
	The Faraday effect, also known as Faraday rotation, is a magneto-optical phenomenon in which the plane of polarization of linearly polarized light rotates as it passes through a transparent material under the influence of a magnetic field, or magnetization, parallel to the direction of light propagation. On the other hand, the inverse Faraday effect as defined by Pitaevskii\cite{pitaevskii_electric_1961} was expressed in transparent materials that were non-magnetic, or not submitted to any external magnetic field. In such a setup, circularly polarized light would induce a magnetization which direction depends on its helicity and which magnitude is proportional to its intensity. While this phenomenon has been measured a while ago\cite{van_der_ziel_optically-induced_1965}, it does not appear to directly correspond to the reciprocal effect of the Faraday rotation. Indeed, such an effect would have to take place in a similar setup as the Faraday rotation, \ie\ within a transparent material under the influence of a magnetic field, or magnetization, in which a linearly polarized light propagates.
	On the other hand, and as mentioned in Sec. \ref{sec:iso_mag}, linearly polarized light generates a second harmonics magnetization density when traveling through such a media, thus being the direct ``inverse" effect of the Faraday rotation. This is further evidenced by the fact that both effects are proportional to $\varepsilon_{xy}$. According to Onsager reciprocal relations, this effect would therefore be the inverse of the Faraday rotation, \ie\ the IFE.

	\section{Conclusion}
	
	Within the framework of linear response theory and the dipole approximation, we derived the general expression of the torque induced by the spin part of light on the electronic system. We then showed that this light-induced torque can readily be expressed in terms of a measurable quantity: the dielectric tensor. When considering time-independent dielectric properties, this torque can equivalently be expressed as an induced magnetization per unit of fluence, thus pinpointing the difference between this phenomenon and the inverse Faraday effect as defined by Pitaevskii\cite{pitaevskii_electric_1961}, being a light-induced magnetization per unit of intensity. This key difference means that, contrary to the inverse Faraday effect, the light-induced torque generates an orbital magnetization that remains, even after the light is gone.
	We showed that, in dissipative media, the absorption of circularly polarized light indeed changes the electronic angular momentum by $\hbar$ per photon. Moreover, and in agreement with the work of Beth\cite{beth_mechanical_1936}, orbital angular momentum can be absorbed even in transparent materials, as long as the spatial symmetry of the material is not isotropic.
	When occurring in magnetic media, it is yet not clear how this absorbed orbital angular momentum --and thus magnetization-- interacts with the existing one. Rather than being considered to act as a so-called effective "opto-magnetic field" as usually assumed in the case of the IFE\cite{hertel_theory_2006, berritta_ab_2016}, such an induced magnetization should more naturally be considered to interact with the existing one through an exchange. While assessing the magnitude of the latter is a challenging task, such a phenomenon could thus drive the various light-induced helicity-dependent phenomena such as the all-optical helicity-dependent switching\cite{stanciu_all-optical_2007,mangin_engineered_2014, lambert_all-optical_2014}, the helicity-dependent domain-wall motion\cite{quessab_helicity-dependent_2018}, and the helicity-dependent demagnetization\cite{tsema_helicity_2016, quessab_resolving_2019}, even long after the light is gone.
	Finally, mutualizing pump-probe techniques to assess the variations of the dielectric response through time with results from this work would allow experimentalists to explicitly compute the amount of orbital angular momentum absorbed from the pump.

	\section{Acknowledgments}
	
	Philippe Scheid is grateful for the many fruitful discussions with B. Herzog, C. Chatelain, A. Bergman and I. P. Miranda.\\
	This work is supported by the French National Research Agency ANR for  ANR-20-CE09-0013 UFO, ANR-15-IDEX-04-LUE (“Lorraine Université d’Excellence”) and through the France 2030 government grants EMCOM (ANR-22-PEEL-0009).

\section{Annexes}

\subsection{Fourier transform}

We define the Fourier transform of a time-dependent function, $f$ as:

\begin{equation}
	\tilde{f}(\omega) = \int_{- \infty}^{\infty} f(t) e^{i \omega t} dt,
\end{equation}

where $\omega$ is the pulsation of the light, and such that the inverse transform is:

\begin{equation}
	f(t) = \frac{1}{2 \pi}\int_{- \infty}^{\infty} \tilde{f}(\omega) e^{- i \omega t} d\omega.
\end{equation}

\subsection{Work done by an electric field}

The density of work per unit of time exerted by the electric field on the electronic system writes as:

\begin{equation}\label{eq:power_ind_density}
	P_\text{abs} (t) = \rho_e \frac{d}{dt} \delta \braket{\hat{\bm{r}}}_\text{ind}(t) \cdot -e \bm{E} (t)
\end{equation}

\begin{widetext}
	\begin{equation}\label{}
		P_\text{abs} (t) = \frac{\varepsilon_0}{(2 \pi)^2} \int_{-\infty}^{+\infty} d\omega \int_{-\infty}^{+\infty} d\omega' \left( - i \omega \tilde{\chi}_{\bm{p}}(\omega) \cdot \tilde{\bm{E}}(\omega) \right) \cdot \tilde{\bm{E}}(\omega') e^{- i (\omega + \omega') t}
	\end{equation}
\end{widetext}

\begin{equation}\label{eq:power}
	\begin{split}
		P_\text{abs} (t) &= \frac{1}{2} \Omega \varepsilon_0 E^2_0 \big[
		\mathfrak{Im}\left[e^{-i 2 \Omega t} \left(\tilde{\chi}_{\bm{p}}(\Omega)\cdot \bm{u}^*\right) \cdot \bm{u}^*\right]\\
		& + \mathfrak{Im}\left[\left(\tilde{\chi}_{\bm{p}}(\Omega) \cdot \bm{u}^*\right) \cdot \bm{u} \right] \big].
	\end{split}
\end{equation}

Following the same proceeding as in Sec. \ref{sec:application}, Eq. \ref{eq:power} can be rewritten as an absorbed energy, $E_\text{abs}$, per unit of fluence such as:

\begin{equation}\label{eq:energy_abs_per_fluence}
	\begin{split}
		\frac{d E_\text{abs}}{d F} (t) &= \frac{\Omega}{c} \big[
		\mathfrak{Im}\left[e^{-i 2 \Omega t} \left(\tilde{\chi}_{\bm{p}}(\Omega)\cdot \bm{u}^*\right) \cdot \bm{u}^*\right]\\
		& + \mathfrak{Im}\left[\left(\tilde{\chi}_{\bm{p}}(\Omega) \cdot \bm{u}^*\right) \cdot \bm{u} \right] \big].
	\end{split}
\end{equation}

\subsection{Number of absorbed photons in an isotropic, dissipative medium} \label{an:number_abs_photons}

In the case of an isotropic medium, the dielectric tensor writes as in Eq. \ref{eq:iso_dielectric} and the density of absorbed energy per unit of fluence writes as:

\begin{equation}
	\begin{split}
		\frac{d E_\text{abs}}{d F} = \frac{\Omega}{c} \varepsilon''_{xx} \\
		+ \frac{\Omega}{c} \varepsilon''_{xx} \left( \cos^2 \left(\theta\right) + \sin^2 \left(\theta\right) \cos \left(2 \phi\right) \right) \cos \left(2\Omega t\right)\\
		- \frac{\Omega}{c} \varepsilon''_{xx} \sin^2 \left(\theta\right) \sin \left(2 \phi\right) \sin \left(2\Omega t\right)\\
		+ \frac{\Omega}{c} \left(1 - \varepsilon'_{xx}\right) \sin^2 \left(\theta\right) \sin \left(2 \phi\right) \cos \left(2\Omega t\right)\\
		- \frac{\Omega}{c} \left(1 - \varepsilon'_{xx}\right) \sin^2 \left(\theta\right) \sin^2 \left(\phi\right) \sin \left(2\Omega t\right)\\
	\end{split}
\end{equation}

For any polarization of the light, the absorptive part thus writes as:

\begin{equation}
	\frac{d E_\text{abs}^\text{stat.}}{d F} = \frac{\Omega}{c} \varepsilon''_{xx} 
\end{equation}

On the other hand, the second harmonic generation cancels for circular polarizations, while it writes as:

\begin{equation}
	\frac{d E_\text{abs}^\text{2\textsuperscript{nd}h.}}{d F} = \frac{\Omega}{c} \left( \varepsilon''_{xx}  \cos \left(2 \Omega t \right) + \left( 1 - \varepsilon'_{xx} \sin \left(2 \Omega t\right) \right) \right)
\end{equation}

for linearly polarized light.

In the case of a circularly polarized light, the number of absorbed photons per unit of fluence therefore writes as:

\begin{equation}\label{eq:n_abs_photon}
	\frac{d n}{d F} = \frac{1}{c \hbar} \varepsilon''_{xx}
\end{equation}

\bibliography{bibliography.bib}

\end{document}